\documentstyle[epsf]{mn}
\newcommand{\etal}{{et al}\/.}
\begin{document}
\title[Nuclei in 3CRR radio galaxies]{Radio, optical and X-ray nuclei
in nearby 3CRR radio galaxies}
\author[M.J.~Hardcastle \& D.M. Worrall]{M.J.\ Hardcastle and
D.M.\ Worrall \\
Department of Physics, University of Bristol, Tyndall Avenue,
Bristol BS8 1TL\\}
\maketitle
\begin{abstract}
{\it HST} observations have shown that low-redshift 3CR radio galaxies
often exhibit a point-like optical component positionally coincident
with the GHz-frequency radio core. In this paper we discuss the
correlation between the luminosities of the radio, optical and X-ray
cores in these objects, and argue that all three components have a
common origin at the base of the relativistic jets.  In unified
models, FRI radio galaxies should appear as dimmed, redshifted
versions of BL Lac objects. We show that such models are consistent
with the spectral energy distributions of the radio galaxies only if
the nuclear X-ray emission in radio galaxies is inverse-Compton in
origin.
\end{abstract}

\section{Introduction}

On sub-arcsecond scales, the central radio components, or `cores', of
radio galaxies are understood, as a result of VLBI observations,
to be the unresolved bases of the relativistic jets that transport
energy to the lobes, seen in partly self-absorbed synchrotron
radiation. Superluminal motion is seen in some sources, and the
supposed unification of radio galaxies with BL Lac objects and quasars (e.g.\
Urry \& Padovani 1995) requires the cores to be relativistically
beamed with Lorentz factors $\ga 5$.  Hardcastle \& Worrall (1999) and
Canosa \etal\ (1999) have found that the nuclear soft-X-ray emission
in radio galaxies is well correlated with the core radio emission.
This finding implies that the X-ray emission is also relativistically beamed
and so must originate in the jet, and is qualitatively
consistent with models in which the strong X-ray emission of BL Lac
objects is largely a result of relativistic boosting.

A snapshot survey of 3CR radio sources with the {\it HST} Wide Field
and Planetary Camera 2 (WFPC2) (De Koff \etal\ 1996) has detected
unresolved optical nuclear components in a large number of objects,
particularly at low redshift (Martel \etal\ 1999), where, because of
the luminosity-redshift correlation in the 3CR sample, most radio
sources are of type FRI rather than FRII (Fanaroff \& Riley
1974). Chiaberge, Capetti \& Celotti (1999) find that the optical
luminosities of the FRI radio galaxies are correlated with the
luminosities of their 5-GHz radio cores, which, by the same argument
as used by Hardcastle \& Worrall for the X-ray emission, implies a
jet-related origin for the nuclear optical emission. Capetti \&
Celotti (1999) compare the optical nuclear luminosities of five of the
detected FRI radio galaxies with those of BL Lac objects matched in
isotropic properties and find support for unification in the optical
assuming relativistic jets with Lorentz factors in the range 5 to
10. In this paper we examine the radio, optical and X-ray nuclear
fluxes and luminosities of a sample of low-redshift radio galaxies in
order to explore the X-ray and optical emission mechanisms.

Throughout the paper we adopt a cosmology with $H_0 = 50$ km
s$^{-1}$ Mpc$^{-1}$. Spectral index $\alpha$ is defined in the
sense that flux density is proportional to $\nu^{-\alpha}$.

\section{Observations and data reduction}

We selected the 27 radio galaxies with $z<0.06$ from the 3CRR sample
(Laing, Riley \& Longair 1983, Laing \& Riley, in prep.); the redshift
limit was chosen to ensure good spatial resolution, so that nuclei
could adequately be separated from the dusty discs that surround
them. The status of the archival WFPC2 {\it HST} observations of these
objects is tabulated in Table \ref{ss}. The majority of the
observations were made as part of the snapshot survey, and so are
short (280 s) and use the Planetary Camera (PC) with the F702W filter,
but we have used longer observations in this or other broad-band red
filters where they exist in the archives. Bias subtraction and
flat-field corrections had been applied automatically as part of the
standard pipeline, and we performed cosmic ray rejection using the
{\sc crrej} program in {\sc iraf}.

Almost every observed source (17/20) in this sample has an optical
nucleus which is unresolved with the {\it HST} (i.e., has a linear
size less than $\sim 50$ pc) , and the majority (13) of these also show
disc-like dust features. The two broad-line objects in the sample,
3C\,382 and 3C\,390.3, show atypically strong nuclei, as would be
expected from their ground-based classification as N-galaxies.

For the 17 sources with detected point-like nuclei we used {\sc iraf}
to carry out small-aperture photometry on the nucleus. Source regions
were typically only 3--5 PC pixels (1 pixel is 46 milliarcsec), and
backgrounds were taken from an annulus around the source close to the
nucleus (up to 10 pixels) to minimise the contribution from the
stellar continuum to the net core emission.  The resulting
measurements were corrected for flux lying outside the extraction
regions using the aperture corrections given by
Holtzman \etal\ (1995), and then converted to flux densities at the
central frequency of the filter ($4.3 \times 10^{14}$ Hz for F702W,
$3.8 \times 10^{14}$ Hz for F791W and F814W) using the {\sc synphot}
package in {\sc iraf}, assuming a power-law spectrum with $\alpha =
0.8$ (the results are insensitive to the assumed spectrum). Flux
densities are tabulated in Table \ref{fluxes}, together with
corresponding radio and X-ray values.

Our analysis is independent of that of Chiaberge \etal\ (1999), though
their sample contains a large number of objects (12 out of the 17 in
Table \ref{fluxes}) in common with ours. Comparing the flux densities
tabulated in their table 3 with our measurements for the
overlapping objects, we find agreement to better than a factor of 2 in
almost all cases, which is reasonable considering the
uncertainties in background subtraction.

\section{Results}

\subsection{Radio-optical correlation}

\begin{figure}
\caption{The radio-optical luminosity-luminosity correlation for $z<0.06$ 3CRR sources}
\label{ro}
\epsfxsize \linewidth
\epsfbox{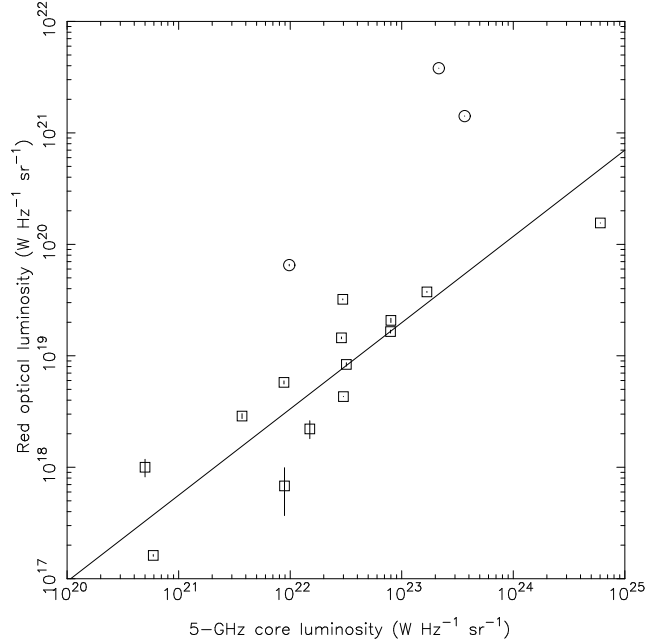}
\begin{minipage}{\linewidth}
\vskip 5pt Squares indicate FRI objects, circles are FRIIs. The two
FRIIs at the top right of the plot are the broad-line objects 3C\,382
and 3C\,390.3, and the other is 3C\,192. The FRI to the far right is
3C\,84. The optical fluxes of the brightest objects (3C\,84, 3C\,382
and 3C\,390.3) are affected by saturation of the {\it HST} CCDs, and
they are all known to be variable in the radio and X-ray, rendering
their positions the most uncertain. The solid line shows the results
of a least-squares linear regression on the data for 14 FRI sources.
\end{minipage}
\end{figure}

The 5-GHz radio versus optical correlation for the 17 sources with
point-like optical cores is shown in Fig.\ \ref{ro}. The
correlation is strong over several orders of magnitude of luminosity,
extending further than that of Chiaberge \etal\ (1999) primarily
because of the inclusion of the luminous broad-line objects 3C\,382
and 3C\,390.3. The optical and radio flux densities are also well
correlated, and a partial Kendall's $\tau$ test shows that the
luminosity-luminosity correlation is not induced by a common
correlation with redshift (at the 95 per cent confidence level).

The one narrow-line FRII in the sample, 3C\,192, lies some way off the
correlation. This may be an artefact of the lower resolution of the
{\it HST} observations (Table \ref{ss}), or it may indicate a
difference between FRIs and FRIIs.

All three objects which do not have a measured point-like optical
nucleus in the {\it HST} observations do not have a nucleus for
reasons which are likely to be unrelated to their true optical nuclear
flux (see notes to Table \ref{ss}). The sub-sample of 17 sources that we
consider is therefore not biased with respect to the optical flux, and
the three missing objects would not be expected to alter the correlation
of Fig.\ \ref{ro}.

\subsection{Optical-X-ray correlation}

\begin{figure}
\caption{The X-ray-optical luminosity-luminosity correlation for $z<0.06$ 3CRR sources}
\label{rx}
\epsfxsize \linewidth
\epsfbox{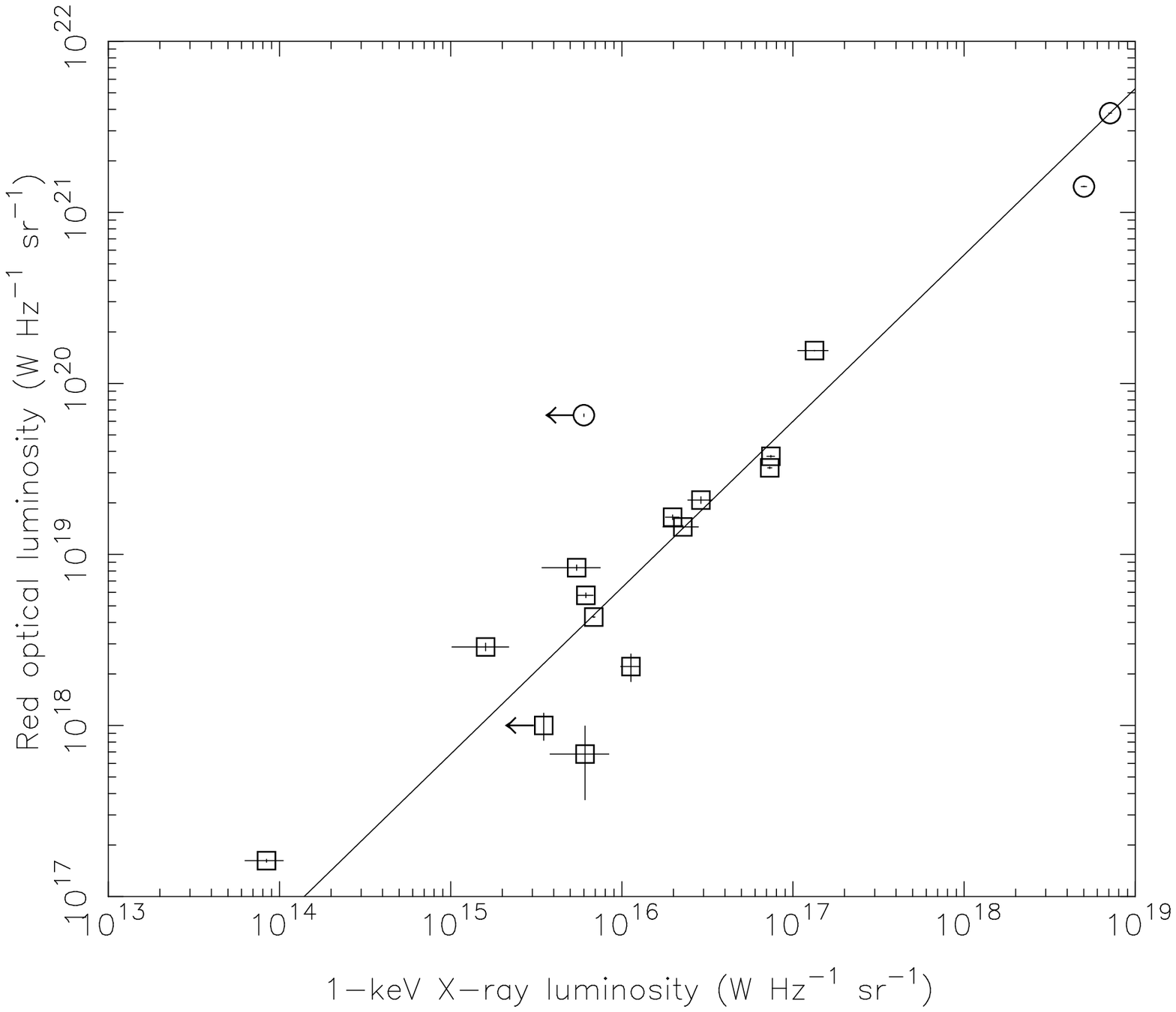}
\begin{minipage}{\linewidth}
\vskip 5pt
Symbols and notes as for Fig.\ \ref{ro}.
\end{minipage}
\end{figure}

The 17 sources with point-like optical cores were all targets for {\it
ROSAT} pointed observations, and X-ray flux densities for the cores
(almost all of which were detected) are given by Hardcastle \& Worrall
(1999) and tabulated in Table \ref{fluxes}.  The X-ray versus radio
correlation (Fig.\ \ref{rx}) is at least as good as that seen in the
radio-optical plot (Fig.\ \ref{ro}), and the scatter about a
regression line appears to be considerably less, suggesting that there
may be a closer relationship between the X-ray and optical
emission. Again, 3C\,192 is an outlier.

\subsection{Colour-colour plots}

\begin{figure}
\caption{Two-point radio-to-optical and optical-to-X-ray spectral
indices}
\label{alpha}
\epsfxsize \linewidth
\epsfbox{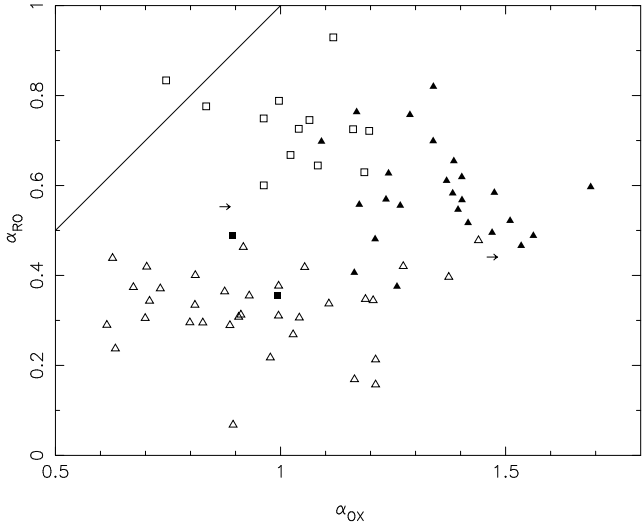}
\vskip 5pt
\begin{minipage}{\linewidth}
Boxes denote radio galaxies, and filled boxes are the broad-line radio
galaxies. Arrows denote lower limits due to X-ray
non-detections. Plotted with our data points are the BL Lacs from the
sample of Fossati \etal\ (1998). Here filled triangles are
radio-selected BL Lac objects from the 1-Jy sample and open triangles are
X-ray selected BL Lacs from the Slew survey. The line indicates the locus of
spectra that can be described as single power laws.
\end{minipage}
\end{figure}

Rest-frame
radio-to-optical and optical-to-X-ray spectral indices ($\alpha_{\rm
RO}$ and $\alpha_{\rm OX}$, respectively) are given in Table
\ref{fluxes} and plotted in Fig.\ \ref{alpha}. Also plotted for comparison
are the corresponding spectral indices for the X-ray-observed BL Lac objects
with measured redshifts in the sample of Fossati \etal\
(1998). K-corrections were made to all the data points using
$\alpha_{\rm R} = 0$, $\alpha_{\rm O} = 0.8$ and $\alpha_{\rm X} =
0.8$, and the BL Lac optical fluxes are also `corrected' from the V-band
measurements tabulated by Fossatti \etal\ to our observing wavelength
of $\sim 7000$ \AA .

We might expect from unified models that the cores of the FRI radio
galaxies would occupy a similar region of the $\alpha_{\rm
RO}$ -- $\alpha_{\rm OX}$ plane to radio-selected BL Lac objects
(RBLs). In fact, Fig.\ \ref{alpha} shows that they occupy a region
which overlaps with the RBLs, and is clearly distinguishable from the
region occupied by X-ray selected BL Lacs (XBLs); the only two
objects to overlap with the XBLs are the broad-line radio galaxies 3C\,382
and 3C\,390.3. However, there is a clearly distinguishable difference
between the radio galaxies and RBLs, in the sense that radio galaxies
tend to have steeper $\alpha_{\rm RO}$ and flatter $\alpha_{\rm OX}$
than RBLs. Since the distribution of $\alpha_{\rm RX}$ of FRI
radio galaxies, both in this sample and in 3CRR as a whole (Hardcastle
\& Worrall 1999), is very similar to that of RBLs, this result at first
sight implies that radio galaxies tend to have fainter optical cores
than we would expect from the behaviour of BL Lacs. We return to this
point below.

\section{Discussion}

\subsection{FRII objects}

The only narrow-line FRII object in the sample is 3C\,192, which is an
outlier on all the correlations involving optical data, since its
optical flux is over-bright relative to its radio or X-ray flux
density. There is likely to be a significant amount of line
contamination in the {\it HST} flux of this object, and it was
observed with the WFC rather than the PC, reducing our ability to
remove background emission. Nevertheless, it is possible that the
anomalous behaviour of 3C\,192 is an indication of a difference
between FRI and FRII objects. No such difference is apparent in the
X-ray and radio data alone (Hardcastle \& Worrall 1999).

The two broad-line objects, also FRIIs, have colours which are
significantly different from those of the rest of the sample. Both
objects are affected by saturation of the PC CCDs, but this effect
means that we are underestimating the optical fluxes, and they may be
even more extreme outliers on Fig.\ \ref{alpha}. However, both objects
are known to be variable in the radio and X-ray, and the data we have
used are of varying vintage, so it is dangerous to draw conclusions
concerning broad-line FRII galaxies based on Fig.\ \ref{alpha}.

\subsection{The origins of the optical and X-ray emission}

The correlation in Fig.\ \ref{ro} implies that the optical emission
originates in the jet. The most likely emission mechanism is then
synchrotron radiation, particularly as optical synchrotron emission is
seen from the kiloparsec-scale jets of several radio galaxies
(e.g. Martel \etal\ 1998 and references therein), including 3C\,66B,
3C\,264 and M87 in our sample.

Several factors make it difficult to constrain the process responsible
for the X-radiation using the radio and optical emission. We know from
the flat radio spectra of the cores that they are self-absorbed at cm
wavelengths, so we can draw no strong conclusions on the origins of
the optical and X-ray emission from the generally convex shapes of
the radio-optical-X-ray spectra; there are few observations of cores
of radio galaxies at radio frequencies high enough to avoid self-absorption.

A well-determined core optical spectrum could provide strong clues
about the mechanism of the X-rays emission. A second optical colour,
in the F547M or F555W filters, is available for a few nuclei in our
sample, and in the majority of cases the derived optical spectral
index is steeper than the optical-to-X-ray spectral index, which would
na\"\i vely suggest that the X-ray and optical emission cannot both be
synchrotron emission from a single population of electrons.  However,
even modest amounts of intrinsic or Galactic reddening can make a
substantial difference to the measured optical spectral index (e.g. Ho
1999). Reddening will be significant if the geometry is such that part
of a dusty disc lies along the line of sight to the nucleus. In the
case of NGC\,6251, Ferrarese \& Ford (1999) have shown that the
reddening inferred from a comparison of the obscured and unobscured
regions of the galaxy is consistent with the intrinsic $N_H$ inferred
from X-ray absorption by Birkinshaw \& Worrall (1993), so that the
dusty disc may be obscuring both the optical and X-ray nuclear
emission. 3C\,264, where the plane of the disc appears to be almost
perpendicular to the line of sight, has the flattest measured optical
spectral index in the sample, and so may be the least affected by
reddening, consistent with this picture.

Therefore the present data cannot rule out a synchrotron origin for
the X-ray emission, when the possible effects of intrinsic reddening
are taken into account. If the X-ray cores were synchrotron, it would
go some way towards explaining the tight relationship between the
optical and X-ray nuclear luminosities. However, an inverse-Compton
origin for the X-ray nuclei may be required by unified models, as
discussed in the next section. Further {\it HST} observations are
needed to measure the reddening and the intrinsic optical spectral
index in a large sample of radio galaxies, if broad-band spectra are
to constrain the X-ray emission process.

\subsection{Contraints from unified models}

We can combine the X-ray, optical and radio data for FRI radio galaxies
to provide a new test of the standard BL-Lac/FRI unified model, in which
relativistic beaming is important in the nuclei and BL Lac objects are
FRIs seen at small angles to the line of sight. In a such model, the
spectral energy distributions (SEDs) of FRIs should be dimmed,
redshifted versions of those of RBLs. We can test this model by
investigating the regions that the two populations occupy in the
$\alpha_{\rm RO}$ -- $\alpha_{\rm OX}$ plane of Fig.\ \ref{alpha}.

If BL Lac spectra were one-component power laws extending from the
radio to the X-ray, we would expect that the FRIs and BL Lacs would
populate similar regions of the $\alpha_{\rm RO}$ -- $\alpha_{\rm OX}$
plane. In fact, the SEDs of BL Lac objects between radio and X-ray bands
are often well modelled by a smoothly curved function (e.g. a
parabola; Landau \etal\ 1986, Sambruna \etal\ 1996).  Redshifting and
dimming such a spectrum to produce the expected spectrum of an FRI radio
galaxy should result in both $\alpha_{\rm RO}$ and $\alpha_{\rm OX}$
being somewhat steeper in radio galaxies than in RBLs, which is not
observed. Instead, $\alpha_{\rm OX}$ is flatter in radio
galaxies. (The effect on the SED from cosmological redshift, given
that the RBLs of Fig.\ \ref{alpha} are more distant than our sample of
FRIs, is small compared to that expected from relativistic beaming.)

The nuclear X-ray emission from the FRIs has been
separated from the thermal emission from their host groups or clusters
(Hardcastle \& Worrall 1999), and we are confident that we are not
overestimating the nuclear X-ray emission significantly in these
sources. In contrast, the discrepancy between the $\alpha_{\rm OX}$
distributions for FRIs and BL Lac objects may be underestimated as a
result of a thermal contribution to the X-ray fluxes of the RBLs,
which can be significant at the $\sim 10$ per cent level (e.g.\
Hardcastle, Worrall \& Birkinshaw 1999).

Reddening in the optical nuclei observed with {\it HST}, as a result
of absorption by the dusty disc seen in the {\it HST} images, may have
an effect. The FRIs can be made to have a very similar distribution of
$\alpha_{\rm RO}$ to the RBLs if it is assumed that obscuration causes
us to underestimate the red optical fluxes by a factor $\sim 2$ to $3$
in the radio galaxies only. However, such a high degree of reddening
would imply a typical $A_{\rm V} \sim 2$, and therefore, for typical
gas/dust ratios, an intrinsic column density of order $5 \times
10^{21}$ cm$^{-2}$. Such column densities are somewhat higher than is
observed for the soft X-ray cores of well-studied FRIs
(e.g. Birkinshaw \& Worrall 1993; Worrall \& Birkinshaw 1994). More
importantly, they would mean that we are underestimating the
unabsorbed 1-keV X-ray flux densities of the FRIs (which are
calculated on the assumption of galactic absorption) by a similar
factor. Thus $\alpha_{\rm OX}$ is not strongly affected by reddening
at this level, and we cannot account for the flat $\alpha_{\rm OX}$ in
radio galaxies in this way.

So the relatively flat $\alpha_{\rm OX}$ in FRIs appears to be
real. If unified models are correct, this observation gives us a
strong reason to believe that the X-ray nuclei of FRIs are dominated
by inverse-Compton emission. To make the data consistent with unified
models, it must be the case that the $\alpha_{\rm OX}$ of FRIs is
observed to be relatively flat because a new component of the RBL
spectrum, not well modelled by the simple parabolae of Landau \etal\
(1986), is dominant at X-ray energies in the FRIs but largely
blueshifted out of the observed band in RBLs. In BL Lacs, an
inverse-Compton component of the emission has long been expected to
dominate at high energies, and this component is believed to be
responsible for the second peak in the SED seen in $\gamma$-rays at
MeV--GeV energies (e.g. Fosatti \etal\ 1998); discrepancies between
the simple parabolic models and the data, and the flat spectra seen in
the X-ray in some BL Lacs, can also be attributed to this component
(Worrall 1989; Sambruna \etal\ 1996).

We therefore favour an inverse-Compton model for the nuclear
X-ray emission seen in these nearby radio galaxies. If the X-ray nuclei of
FRIs are indeed dominated by inverse-Compton emission, then we expect them
to have flat soft X-ray spectra, a prediction that will be testable
with forthcoming {\it Chandra} observations.

\section*{Acknowledgements}
This work was supported by PPARC grant GR/K98582.
\begin{table}
\caption{Archival {\it HST} data for 3CRR radio galaxies with $z<0.06$}
\label{ss}
\begin{tabular}{llrr}
Source&Obs. status&Filter&Duration (s)\\
\hline
3C\,31&OK&F702W&280\\
3C\,33&Not observed\\
3C\,66B&OK&F702W&280\\
3C\,76.1&Not observed\\
3C\,83.1B&Diffraction spike&F702W&280\\
3C\,84&Saturated&F702W&560\\
3C\,98&Mispointed\\
DA\,240&Not observed\\
3C\,192&WFC observations&F702W&280\\
4C\,73.08&Not observed\\
3C\,264&OK&F702W&280\\
3C\,272.1 (M84)&OK&F702W&280\\
3C\,274 (M87)&OK&F814W&2430\\
3C\,293&Disturbed&F702W&280\\
3C\,296&OK&F702W&280\\
3C\,305&Extended&F702W&560\\
3C\,310&OK&F702W&280\\
NGC\,6109&Not observed\\
3C\,338&OK&F702W&280\\
NGC\,6251&OK&F814W&1000\\
3C\,382&Saturated&F702W&280\\
3C\,386&Foreground star&F702W&280\\
3C\,390.3&Saturated&F702W&280\\
3C\,442A&OK&F791W&750\\
3C\,449&OK&F702W&280\\
NGC\,7385&Not observed\\
3C\,465&OK&F702W&280\\
\end{tabular}
\begin{minipage}{\linewidth}
\vskip 10pt Notes: Sources with `saturated' optical nuclei yield only
a lower limit on their optical flux densities; 3C\,382 is the worst
affected by this effect. 3C\,83.1B's nucleus lies close to the diffraction
spike from a bright foreground star, so the optical flux density is
uncertain. 3C\,192 was observed in the Wide Field Camera (WFC), so the
nucleus is not well resolved from background emission. 3C\,293 shows
strong, irregular dust features, and it is not clear which
component, if any, is the optical core. 3C\,305's `nucleus' is
extended and off-axis, and probably related to scattering from a more
strongly obscured true nucleus, or affected by line contamination
(Jackson \etal\ 1995). A foreground star lies directly on the line of
sight to the nucleus of 3C\,386 (Lynds 1971), preventing a measurement
of its flux.
\end{minipage}
\end{table}

\begin{table*}
\caption{Radio, optical and X-ray flux densities and spectral indices for the
objects with detected point-like optical nuclei}
\label{fluxes}
\begin{tabular}{lrrrrlrr}
&&\multicolumn{3}{c}{Flux densities at}&&\multicolumn{2}{c}{Spectral indices}\\
Source&$z$&5 GHz&Optical&1 keV&Ref. for&$\alpha_{\rm RO}$&$\alpha_{\rm
OX}$\\
&&(mJy)&($\mu$Jy)&(nJy)&radio\\
\hline
3C\,31&0.0167 &92 &$60 \pm 2 $&$64 \pm 7 $&1&0.64&$1.08$
\\
3C\,66B&0.0215 &182 &$90 \pm 2 $&$140 \pm 30 $&2&0.67&$1.02$
\\
3C\,83.1B&0.0255 &40 &$3 \pm 1 $&$30 \pm 10 $&3&0.83&$0.75$
\\
3C\,84&0.0172 &59600 &$1520 \pm 3 $&$1300 \pm 300 $&4&0.93&$1.12$
\\
3C\,192$\dag$&0.0598 &8 &$51 \pm 0.9 $&$< 5 $&1&0.44&$>1.47$
\\
3C\,264&0.0208 &200 &$213 \pm 2 $&$486 \pm 10 $&5&0.60&$0.96$
\\
3C\,272.1&0.0031 &180 &$49 \pm 0.9 $&$26 \pm 6 $&1&0.72&$1.20$
\\
3C\,274&0.0043 &4000 &$679 \pm 2 $&$1080 \pm 20 $&1&0.79&$1.00$
\\
3C\,296&0.0237 &77 &$10 \pm 2 $&$58 \pm 7 $&1&0.78&$0.83$
\\
3C\,310&0.0540 &80 &$20 \pm 0.9 $&$28 \pm 5 $&1&0.73&$1.04$
\\
3C\,338&0.0298 &105 &$27 \pm 0.9 $&$20 \pm 7 $&1&0.73&$1.16$
\\
NGC\,6251&0.024 &850 &$186 \pm 2 $&$370 \pm 20 $&6&0.75&$0.96$
\\
3C\,382$\dag$&0.0578 &188 &$3180 \pm 6 $&$5950 \pm 100 $&1&0.35&$0.99$
\\
3C\,390.3$\dag$&0.0569 &330 &$1220 \pm 4 $&$4300 \pm 100 $&1&0.49&$0.89$
\\
3C\,442A&0.0263 &2 &$4 \pm 0.7 $&$< 20 $&1&0.55&$>0.88$
\\
3C\,449&0.0171 &37 &$28 \pm 1 $&$20 \pm 6 $&1&0.63&$1.19$
\\
3C\,465&0.0293 &270 &$55 \pm 2 $&$66 \pm 6 $&1&0.75&$1.06$
\\
\end{tabular}
\begin{minipage}{\linewidth}
\vskip 10pt Sources marked with a dagger are classed as FRIIs. Errors
are $1\sigma$ and are statistical only. We have not attempted to
compute the errors in the optical data due to background subtraction
uncertainties. X-ray data are taken from Hardcastle \& Worrall
(1999). Radio references are as follows: (1) Giovannini \etal\ (1988);
(2) Leahy, J\"agers \& Pooley (1986); (3) O'Dea \& Owen (1985); (4)
Noordam \& de Bruyn (1982); (5) Gavazzi, Perola \& Jaffe (1981); (6)
Jones \etal\ (1986).
\end{minipage}
\end{table*}

\end{document}